\documentclass[aps,prd,twocolumn,showpacs,amsmath,amssymb]{revtex4-1}
\usepackage{amsmath} \usepackage{graphicx} \usepackage{subfigure}
\usepackage{epstopdf} \usepackage{color} \usepackage{multirow}
\usepackage{setspace} \usepackage{overpic} \usepackage{amssymb}
\usepackage{ulem}

\usepackage[bookmarksnumbered, pdfstartview=FitH,colorlinks,urlcolor=blue, citecolor=blue,linkcolor=blue] {hyperref}
\usepackage{lineno}
\usepackage{bm}
\usepackage{rotating}
\usepackage{xcolor}
\usepackage{makecell}
\usepackage{mathtext}
\usepackage{mathrsfs}
\usepackage{overpic}
\usepackage[T1]{fontenc}
\usepackage{lmodern}
\usepackage[utf8]{inputenc}
\usepackage{enumitem}

\newcommand{\PreserveBackslash}[1]{\let\temp=\\#1\let\\=\temp}
\newcolumntype{C}[1]{>{\PreserveBackslash\centering}p{#1}}
\newcolumntype{R}[1]{>{\PreserveBackslash\raggedleft}p{#1}}
\newcolumntype{L}[1]{>{\PreserveBackslash\raggedright}p{#1}}
\newcommand{\pp}{\pi^+\pi^-}

\newcommand{\psip}{\psi(3686)}

\newcommand{\jpsi}{J/\psi}

\newcommand{\ppb}{p\bar{p}}
\newcommand{\LLB}{\Lambda\bar{\Lambda}}
\newcommand{\SSB}{\Sigma^{0}\bar{\Sigma}^{0}}
\newcommand{\XXB}{\Xi^-\bar{\Xi}^+}

\newcommand{\etac}{\eta_{c}}
\newcommand{\hc}{h_{c}}

\newcommand{\BESIIIorcid}[1]{\href{https://orcid.org/#1}{\hspace*{0.1em}\raisebox{-0.45ex}{\includegraphics[width=1em]{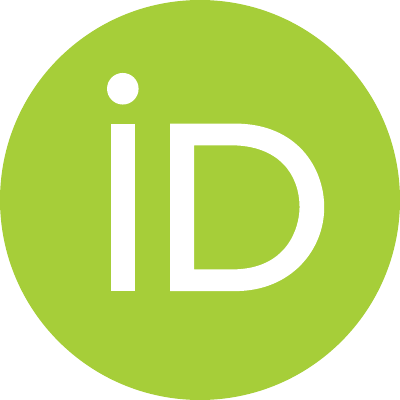}}}}

\let\oldequation\equation
\let\oldendequation\endequation
\renewenvironment{equation}{\linenomathNonumbers\oldequation}{\oldendequation\endlinenomath}

\begin{document}

\graphicspath{{figure/}}
\DeclareGraphicsExtensions{.eps,.png,.ps}

\title{\boldmath Observation of $\eta_c(1S)\to \Sigma^0\bar \Sigma^0$ and search for $h_c(1P)\to \Sigma^0\bar \Sigma^0$ via \texorpdfstring{$\psi$}{psi}(3686) transitions}

\author{
\begin{small}
\begin{center}
M.~Ablikim$^{1}$\BESIIIorcid{0000-0002-3935-619X},
M.~N.~Achasov$^{4,c}$\BESIIIorcid{0000-0002-9400-8622},
P.~Adlarson$^{83}$\BESIIIorcid{0000-0001-6280-3851},
X.~C.~Ai$^{89}$\BESIIIorcid{0000-0003-3856-2415},
C.~S.~Akondi$^{31A,31B}$\BESIIIorcid{0000-0001-6303-5217},
R.~Aliberti$^{39}$\BESIIIorcid{0000-0003-3500-4012},
A.~Amoroso$^{82A,82C}$\BESIIIorcid{0000-0002-3095-8610},
Q.~An$^{78,65,\dagger}$,
Y.~H.~An$^{89}$\BESIIIorcid{0009-0008-3419-0849},
M.~S.~Anderson$^{39}$\BESIIIorcid{0009-0008-1550-2632}
% ...
\\
\vspace{0.2cm}
(BESIII Collaboration)\\
\vspace{0.2cm} {\it
$^{1}$ Institute of High Energy Physics, Beijing 100049, People's Republic of China\\
% ...
$^{4}$ Budker Institute of Nuclear Physics SB RAS (BINP), Novosibirsk 630090, Russia\\
% ...
$^{31}$ INFN Sezione di Ferrara, (A)INFN Sezione di Ferrara, I-44122, Ferrara, Italy; (B)University of Ferrara, I-44122, Ferrara, Italy\\
% ...
$^{39}$ Johannes Gutenberg University of Mainz, Johann-Joachim-Becher-Weg 45, D-55099 Mainz, Germany\\
% ...
$^{65}$ State Key Laboratory of Particle Detection and Electronics, Beijing 100049, Hefei 230026, People's Republic of China\\
% ...
$^{78}$ University of Science and Technology of China, Hefei 230026, People's Republic of China\\
% ...
$^{82}$ University of Turin and INFN, (A)University of Turin, I-10125, Turin, Italy; (B)University of Eastern Piedmont, I-15121, Alessandria, Italy; (C)INFN, I-10125, Turin, Italy\\
$^{83}$ Uppsala University, Box 516, SE-75120 Uppsala, Sweden\\
% ...
$^{89}$ Zhengzhou University, Zhengzhou 450001, People's Republic of China\\
\vspace{0.2cm}
$^{\dagger}$ Deceased\\
% ...
$^{c}$ Also at the Novosibirsk State University, Novosibirsk, 630090, Russia\\
% ...
}
\end{center}
\vspace{0.4cm}
\end{small}
}


\begin{abstract}
Using $(2712.4 \pm 14.3) \times 10^6~\psi(3686)$ events collected by the BESIII detector operating at the BEPCII collider, the hadronic decay $\eta_{c}\to\Sigma^{0}\bar{\Sigma^{0}}$ is observed for the first time via the radiative transition from $\psi(3686)$. It is found that the branching fraction has a significant dependence on the interference pattern between $\etac$ and non-$\etac$ processes. They are determined to be $\displaystyle\mathcal{B}(\etac \to \SSB) = (2.59 \pm 0.14(stat) \pm 0.44(syst)) \times 10^{-3}$ and $(1.18 \pm 0.12(stat) \pm 0.21(syst)) \times 10^{-3}$, for the destructive and constructive interference scenarios, respectively. No significant signal is observed for the decay $h_{c}\to\SSB$ in the hadronic transition $\psi(3686)\to\pi^0\hc$, and an upper limit on its branching fraction is set to be $1.02\times 10^{-4}$ at the 90\% confidence level.
\end{abstract}
\maketitle


\section{Introduction}
The charmonium states below the open-charm threshold have been studied extensively, but questions about their spin-singlet states, for example, the P-wave state $h_{c}$ and the S-wave ground state $\eta_{c}$, remain unresolved. Although the $\eta_{c}$ meson has been investigated for nearly four decades since its discovery~\cite{Ref_etacdis}, most of its decay modes remain unidentified. Even for the known channels, some of their measured branching fractions (BFs) suffer from considerable uncertainties. Early measurements of the $\etac$ BFs in $\jpsi$ and $\psip$ decays did not sufficiently take into account the interference effects between resonant and non-resonant processes. More recent studies, such as the coherent amplitude analysis of $\etac \to 2(\pi^+\pi^-)\eta$~\cite{Ref_yxq}, have partially addressed this limitation.

Experimental studies on exclusive charmonium decays play an important role in testing the perturbative quantum chromodynamics (pQCD). The complexity~of QCD remains unsolved in the charmonium mass region. There are still many contradictions between pQCD calculations and experimental measurements. In particular, the pQCD helicity selection rule ~\cite{Ref_pqrd1,Ref_pqrd2,Ref_pqrd3} is violated in many exclusive charmonium decay processes. Charmonium decay processes with baryon anti-baryon final states, such as $\eta_c/h_c\to B_8{\bar{B}_8}$, where $B_8{\bar{B}_8}$ denotes the octet baryon anti-baryon pair, are supposed to be highly suppressed according to the pQCD helicity selection rule. Theoretical predictions on the BFs of $\eta_c/\hc\to B_8\bar{B}_8$ have been made based on a model consideringthe long distance contribution via charmed hadron loops~\cite{Ref_bb}. Some experimental measurements agree with the theoretical predictions, e.g. $\etac\to\ppb,\LLB$ and $\XXB$; while others do not, e.g. $\etac\to\Sigma^+\Sigma^-$. Meanwhile, its isospin partner $\etac\to\SSB$ has not been observed. A study of this decay channel would allow for a test of isospin invariance. In addition, theoretical predictions for the $h_{c}$ have not been tested due to the lack of experimental information, since its predominant decay mode is the electric dipole (E1) transition of $h_c\to\gamma\eta_c$. Recently, BESIII measured this E1 transition BF to be $(57.66^{+3.62}_{-3.50}\pm 0.58)\%$~\cite{Ref_htoeta}, which is reasonably consistent with some theoretical predictions~\cite{Ref_llh1,Ref_llh2}. This implies that nearly half of the $h_c$ decays proceed via a non-E1 transition. However, the decay modes of $h_c$ remain poorly understood experimentally, even after decades of study. Currently, the sum of measured BFs, excluding the $\gamma\etac$ channel, amounts to only about $6\%$~\cite{Ref_pdg}. Up to now, no $B_8\bar{B}_8$ final state has ever been observed.

In this paper, we present the first observation of $\eta_{c} \to \SSB$ via $\psi(3686)\to \gamma\eta_{c}$ and search for $h_{c} \to \SSB$ via $\psi(3686)\to \pi ^{0} h_{c}$ by analyzing $(2712.4 \pm 14.3) \times 10^6~\psip$ events collected by the BESIII detector~\cite{Ref_total}.

\section{BESIII DETECTOR AND MONTE CARLO SIMULATION}
\label{sec:BES}

The BESIII detector~\cite{Ref_bes3} records symmetric $e^+e^-$ collisions
provided by the BEPCII storage ring~\cite{Yu:IPAC2016-TUYA01}
in the center-of-mass energy range from 1.84 to 4.95 GeV, with a peak luminosity of $1.1 \times 10^{33}\;\text{cm}^{-2}\text{s}^{-1}$
achieved at $\sqrt{s} = 3.773\;\text{GeV}$. Large data samples have been collected in this energy region~\cite{Ablikim:2019hff, EcmsMea, EventFilter}. The cylindrical core of the BESIII detector covers 93\% of the full solid angle and consists of a helium-based multilayer drift chamber~(MDC), a time-of-flight system~(TOF), and a CsI(Tl) electromagnetic calorimeter~(EMC), which are all enclosed in a superconducting solenoidal magnet providing a 1.0~T magnetic field. The solenoid is surrounded by an octagonal flux-return yoke made of steel, interleaved with resistive-plate-counter muon-identification modules. The charged-particle momentum resolution at $1~{\rm GeV}/c$ is $0.5\%$, and the d$E$/d$x$ resolution is $6\%$ for electrons from Bhabha scattering. The EMC measures photon energies with a resolution of $2.5\%$ ($5\%$) at $1$~GeV in the barrel (end cap) region. The time resolution in the plastic scintillator TOF barrel region is 68~ps, while that in the end cap region was 110~ps. The end cap TOF system was upgraded in 2015 using multigap resistive plate chamber technology, providing a time resolution of
60~ps, which benefits 83\% of the data used in this analysis~\cite{Ref_tof1, Ref_tof2, Ref_tof3}.

Monte Carlo (MC) simulated data samples generated with a {\sc geant4}-based~\cite{Ref_geant4} software package, which includes the geometric description of the BESIII detector and its response, are used to determine detection efficiencies and estimate backgrounds. The simulation models the beam energy spread and initial state radiation (ISR) in $e^+e^-$ annihilations using the generator {\sc kkmc}~\cite{Ref_kkmc1, Ref_kkmc2}. The inclusive MC sample includes the production of the $\psi(3686)$ resonance, the ISR production of the $J/\psi$, and the continuum processes incorporated in {\sc kkmc}~\cite{Ref_kkmc1, Ref_kkmc2}. All particle decays are modeled with {\sc evtgen}~\cite{Ref_evtgen1, Ref_evtgen2} using branching fractions either taken from the
Particle Data Group~\cite{Ref_pdg}, when available, or otherwise estimated with {\sc lundcharm}~\cite{Ref_lund1, Ref_lund2}.The exclusive decay of $\psi(3686) \to \gamma\etac$ is generated with the angular distribution of 1 + $\cos^{2}\theta_{\gamma}$, where $\theta_{\gamma}$ is the polar angle of the radiative photon in the rest frame of the $\psi(3686)$~\cite{Ref_polar1, Ref_polar2}. The $\psi(3686)\to \pi ^{0} h_{c}$ and  $\etac/h_{c} \to \SSB$ decays are generated uniformly in phase space (PHSP).

\section{EVENT SELECTION}
\label{sec:selection}
The decays of interest in this analysis are the cascade processes $\psip\to\gamma\etac$ with $\etac\to\SSB$, and $\psip\to\pi^{0}\hc$ with $\hc\to\SSB$, where $\Sigma^{0}\to\gamma\Lambda$($\Lambda\to p\pi^{-}$) and $\bar{\Sigma}^{0}\to\gamma\bar{\Lambda}$($\bar{\Lambda}\to\bar{p}\pi^{+}$). Charged tracks with the polar angle ($\theta$) in the range of $|\cos\theta|<0.93$ are detected in the MDC. Since $\Lambda/\bar{\Lambda}$ contains secondary vertices, no interaction vertex requirement is applied on candidate tracks. Each charged track is identified using particle identification (PID), which combines measurements of the energy deposited in the MDC~(d$E$/d$x$) and the flight time in the TOF to form likelihoods $\mathcal{L}(h)~(h=p,K,\pi)$ for each hadron $h$ hypothesis. Tracks are identified as protons when the proton hypothesis has the greatest likelihood ($\mathcal{L}(p)>\mathcal{L}(K)$ and $\mathcal{L}(p)>\mathcal{L}(\pi)$), while charged kaons and pions are identified by comparing the likelihoods for the kaon and pion hypotheses, $\mathcal{L}(K)>\mathcal{L}(\pi)$ and $\mathcal{L}(\pi)>\mathcal{L}(K)$, respectively. At least four charged particles, $\ppb\pp$, need to be identified.

Photon candidates are reconstructed from neutral showers in the EMC. The deposited energy of each candidate is required to be greater than 25 MeV for the barrel region $(|\cos\theta|<0.80)$ and 50 MeV for the end cap region $(0.86<|\cos\theta|<0.92)$.To suppress electronic noise and showers unrelated to the event, the difference between the EMC time and the event start time is required to be within
[0, 700]\,ns.

For the process $\etac\to\SSB$, it is required that there are at least three photons. A four-constraint (4C) kinematic fit is performed
under the $\psip\to 3\gamma p\bar{p}\pi^{+} \pi^{-}$ hypothesis. If there are multiple combinations of $3\gamma p\bar{p}\pi^{+} \pi^{-}$, the one with the smallest $\chi^2_{\rm 4C}$ is chosen.
Furthermore, $\chi_{\rm 4C}^2<20$ is required, which is obtained by optimizing the figure-of-merit (FOM), defined as $S/\sqrt {S + B}$. Here, $S$ denotes the number of signal MC events scaled with the expected branching fraction, while $B$ denotes the number of background events from the inclusive MC sample.

\begin{figure}[htb]
\begin{center}
\includegraphics[width=.4\textwidth]{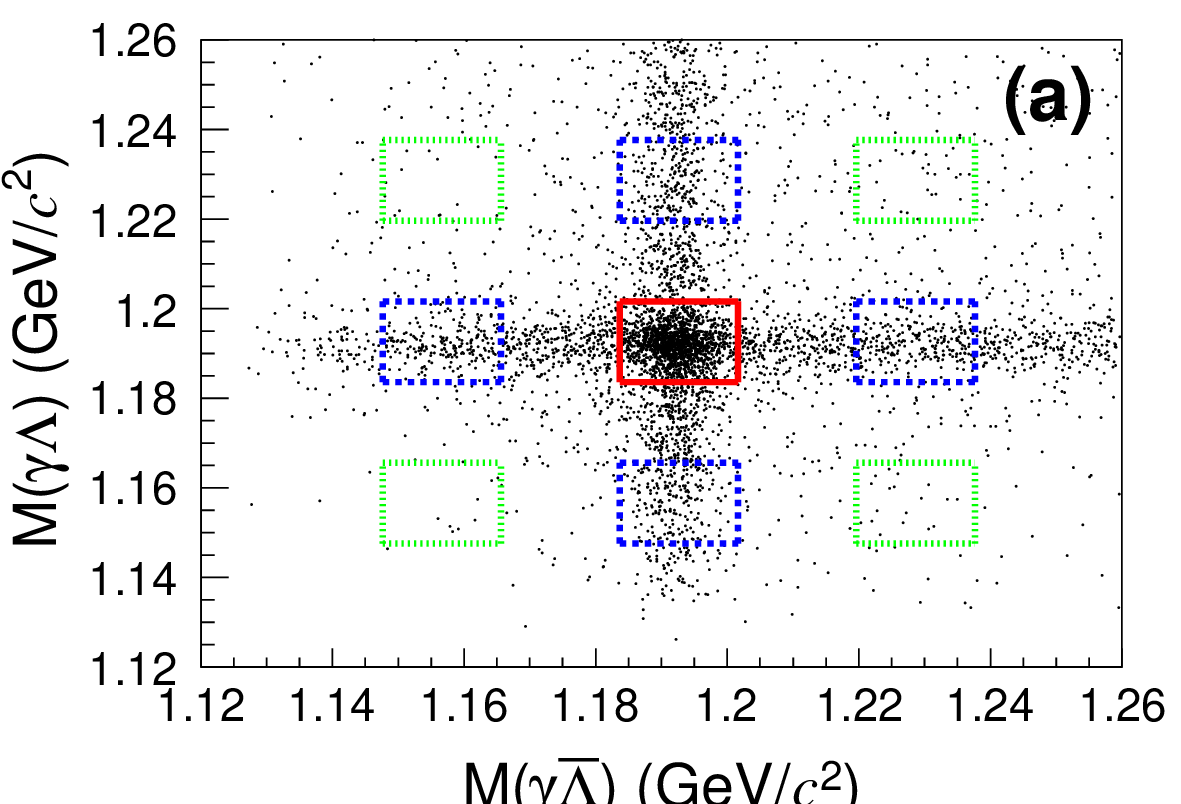}
\includegraphics[width=.4\textwidth]{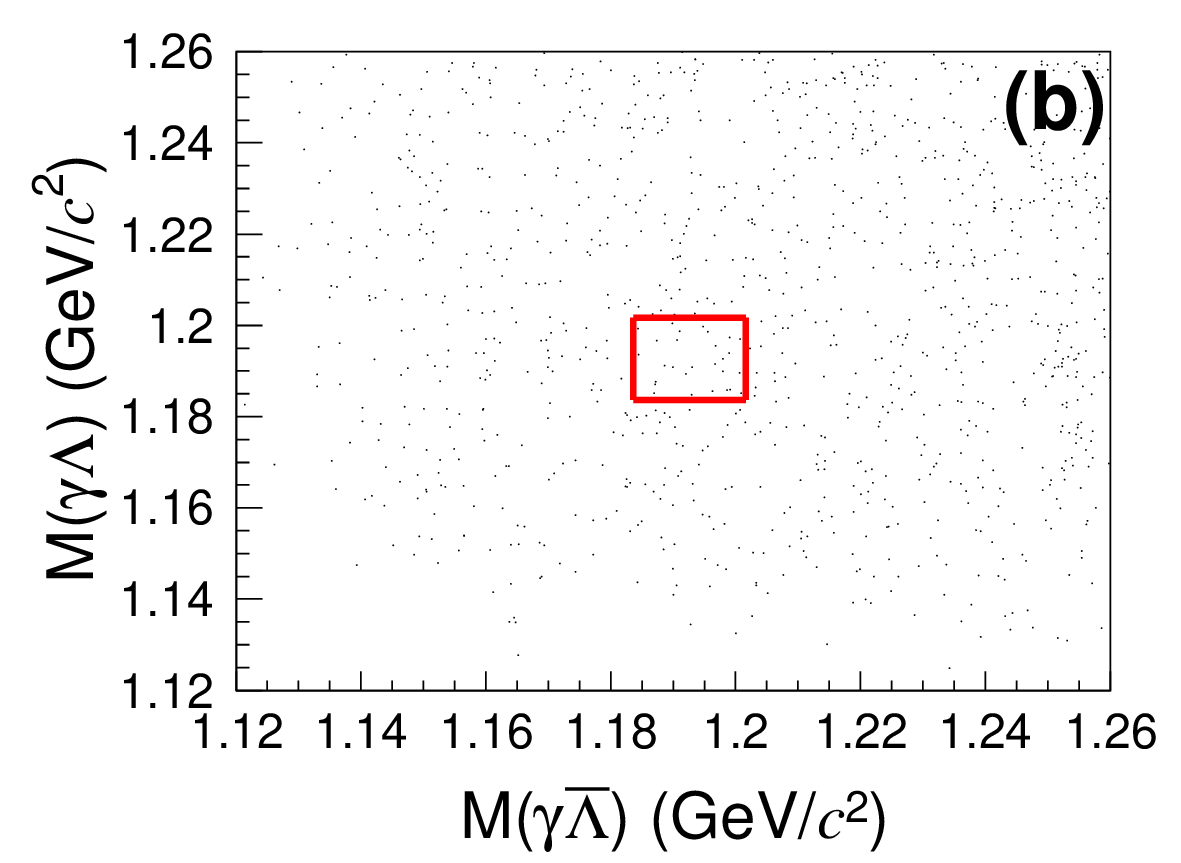}
\caption{The two-dimensional (2D) distribution of $M(\gamma \Lambda)$ versus $M(\gamma\bar{\Lambda})$ for (a)$\etac\to\SSB$ and (b)$\hc\to\SSB$. The red box with solid lines denotes the $\SSB$ signal region, and the blue boxes with dotted lines and green boxes with dashed lines denote the $\SSB$ sideband regions.}
\label{fig:masswinss}
\end{center}
\end{figure}

To reconstruct the $\Lambda/\bar{\Lambda}$, we perform a vertex fit and subsequently a secondary vertex fit to the $p\pi^-/\bar{p}\pi^+$ combination. Furthermore, the decay length of $\Lambda/\bar{\Lambda}$ is required to be greater than 0 to suppress the potential background. If more than one $\Lambda\bar{\Lambda}$ pair candidate survives, the combination that minimizes the mass difference is selected. The mass difference is defined as $\Delta_m  =\sqrt{(M(p\pi^{-})-m_{\Lambda } )^{2}+(M(\bar{p}\pi^{+})-m_{\bar{\Lambda}})^{2}}$, where $m_{\Lambda }$ and $m_{\bar{\Lambda}}$ are the nominal masses of the $\Lambda$ and $\bar{\Lambda}$, respectively. The $\Lambda(\bar{\Lambda})$ signal region is set to be $M(p\pi^{-})(M(\bar{p}\pi^{+}))\in [1.108, 1.124]$ GeV/$c^{2}$. Subsequently, the $\SSB$ pair signal is selected by minimizing mass difference $\Delta_m =\sqrt{(M(\gamma \Lambda)-m_{\Sigma ^{0} })^{2} + (M(\gamma \bar{\Lambda})-m_{\bar{\Sigma}^{0}})^{2}}$, where $m_{\Sigma ^{0}}$ and $m_{\bar{\Sigma ^{0}}}$ are the nominal masses of the $\Sigma ^{0}$ and $\bar{\Sigma^{0}}$, respectively. The $\Sigma ^{0}(\bar{\Sigma}^{0})$ signal region is set to be $M(\gamma \Lambda)(M(\gamma \bar{\Lambda}))\in [1.184, 1.202]$ GeV/$c^{2}$, shown by the red box in Fig.~\ref{fig:masswinss}(a). The eight boxes around the signal region with the same area are taken as sideband regions, as shown in Fig.~\ref{fig:masswinss}. The sideband regions are defined as [1.148, 1.166] GeV/$c^{2}$ and [1.220, 1.238] GeV/$c^{2}$ for $\Sigma ^{0}$ and $\bar{\Sigma}^{0}$. The middle sideband regions, indicated by blue boxes, represent backgrounds with one correct $\Sigma^{0}$ or $\bar{\Sigma}^{0}$ and the other incorrect, denoted as SID1. The corner sideband regions, indicated by green boxes, represent the backgrounds with both incorrect $\Sigma^{0}$ and $\bar{\Sigma}^{0}$, denoted as SID2. The total sideband sample is divided into 125 sub-samples with an equal interval in the $M(\SSB)$ distribution from 2.70 to 3.20 GeV/$c^{2}$. Within each bin, the number of events is obtained by normalizing to the sideband, i.e., $N_{\rm bkg}=\frac{1}{2}\times N_{\rm bkg}^{\rm SID1}-\frac{1}{4}\times N_{\rm bkg}^{\rm SID2}$. Here, $N_{\rm bkg}^{\rm SID1}$ and $N_{\rm bkg}^{\rm SID2}$ denote the numbers of events in the SID1 and SID2 regions, respectively.

For the process $\hc\to\SSB$, it is required that there are at least four photons. A 5C kinematic fit is performed under the $\psip \to 4\gamma p\bar{p}\pi^{+} \pi^{-}$ hypothesis with an extra 1C mass constraint for the accompanying $\pi^0$. If there are multiple combinations of $4\gamma p\bar{p}\pi^{+} \pi^{-}$, the one with the smallest $\chi^2_{\rm 5C}$ is kept for the further analysis. Furthermore, $\chi_{\rm 5C}^2<30$ is required, which is obtained by optimizing the FOM. Figure~\ref{fig:masswinss}(b) shows the distribution of $M(\gamma\Lambda)$ versus $M(\gamma\bar{\Lambda})$.

\section{BACKGROUND ESTIMATION}

The inclusive MC events are employed to investigate the potential background by an event-type analysis tool TopoAna~\cite{Ref_topo}. For the $\psip\to\gamma\etac$ decay mode, no peaking background is found. The fake $\pi^{0}$ background from the wrong photon combination is suppressed by the $\gamma\gamma$ mass window, i.e., $M(\gamma_{1}\gamma_{2})\notin [0.12,0.15]$GeV/$c^{2}$ and $M(\gamma_{1}\gamma_{3})\notin [0.12,0.15]$GeV/$c^{2}$, where $\gamma_{1}$ is the radiative photon from $\psip$ decay and $\gamma_{2}$ and $\gamma_{3}$ from $\SSB$ decay. For the background of missing one photon in the process of $\psip\to\pi^{0}\SSB$, a data-driven method~\cite{Ref_driven} is employed to determine its contribution, by comparing the detection efficiencies between the signal and background channels, and incorporating the observed number of events from the data. The yield of the $\psi(3686) \to \pi^{0}\SSB$ events is obtained by fitting the $M(\gamma\gamma)$ distribution. The corresponding background contribution is fixed according to the data-driven result by fitting the $\etac$ mass region. The $J/\psi$-related background includes two parts: $\psip\to \gamma \chi _{c1,2}  \to \gamma \gamma J/\psi \to\gamma \gamma \SSB$ and $\psip\to \pi^{0}J/\psi \to\gamma \gamma \SSB$. Their contributions are taken into account by the BF normalization.
The continuum background is investigated with the data at $\sqrt{s}=$3.65 GeV. Using the same selection criteria as mentioned above, it is found that almost no events survive in both decay channels. Therefore, the continuum background is negligible.


\section{\label{Sec:BR_determined}The BF Determination }

\subsection{$\etac \to \SSB$}

\begin{figure}[htb]
\begin{center}
\includegraphics[width=.4\textwidth]{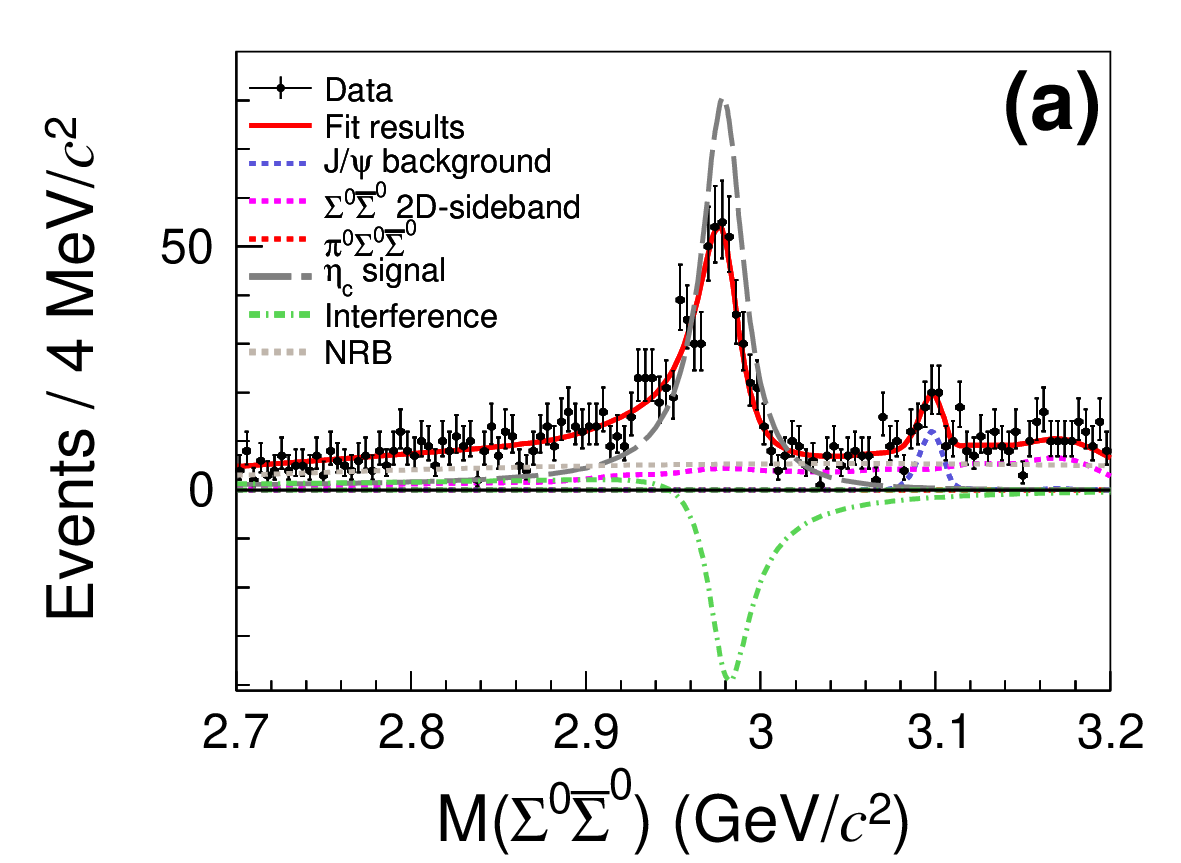}
\includegraphics[width=.4\textwidth]{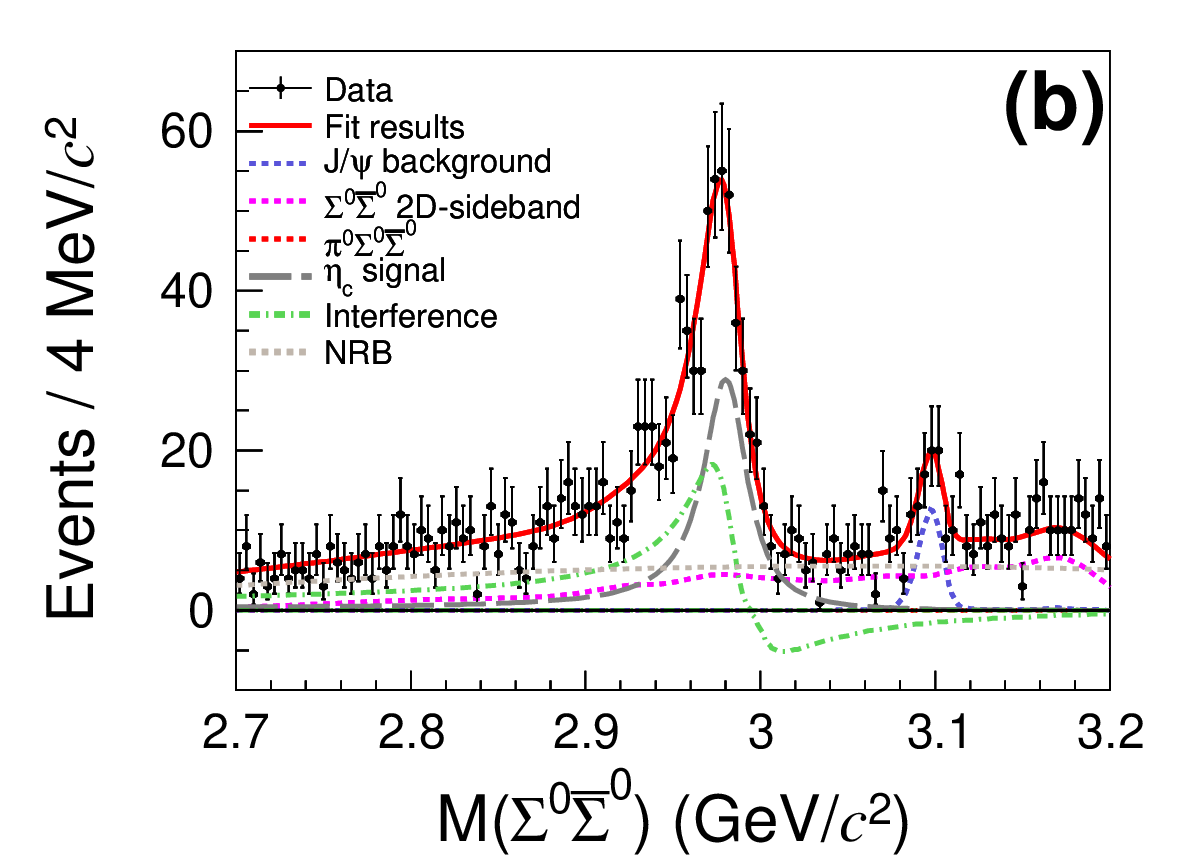}
\caption{The fits to the $M(\Sigma^0\bar \Sigma^0)$ distribution of the accepted candidates for $\eta_c\to \Sigma^0\bar \Sigma^0$ in the data,
with  (a) destructive and (b) constructive interference solutions. The different components are expressed with different line styles and colors.}
\label{1sfit}
\end{center}
\end{figure}

The $\etac$ signal yield is obtained by an unbinned maximum likelihood fit to the $M(\SSB)$ distribution of the accepted candidates for $\etac\to \Sigma^0\bar \Sigma^0$ in the data, as shown in Fig.~\ref{1sfit}. The background includes four components: the $\psi(3686)\to \pi^{0}\SSB$ channel, the non-resonant background~(NRB) from $\psi(3686)\to \gamma\SSB$, the contribution from the $\SSB$ sideband, and the $\jpsi$-related backgrounds. Taking into account the interference between the $\etac$ and the non-resonant background, the probability density function~(PDF) of the fit is expressed as
\begin{equation}
\label{lineshape}
\textstyle{G\otimes|e^{i\phi }E_{\gamma }^{7/2}\sqrt{damp}\text{BW}(m)+ \alpha \text{NRB}|^{2}+ \text{BKG}}.
\end{equation}
Here, $G$ denotes a Gaussian function with free parameters, accounting for the difference in detector resolution between data and MC simulation. The $m$ denotes the $M(\SSB)$ and $\phi$ denotes the interference phase. The energy of the radiative photon of $\psip\to\gamma\etac$ is given by $E_{\gamma} = \frac{m_{\rm \psip}^2 - m^2}{2m_{\rm \psip}}$, where $m_{\psip}$ is the nominal mass of $\psip$. The BW$(m)$ corresponds to a Breit-Wigner function, defined as BW$(m) = \frac{\Gamma}{2\pi} \cdot \frac{1}{(m - m_{\eta_c})^2 + \frac{\Gamma^2}{4}}$, where $m_{\eta_c}$ and $\Gamma$ denote the nominal mass and width of the $\eta_c$~\cite{Ref_pdg}, respectively. The damping function is defined as $damp = \frac{E_0^2}{E_{\gamma}E_0 + (E_{\gamma}-E_0)^2}$, with $E_0 = \frac{m_{\psip}^2 - m_{\eta_c}^2}{2m_{\psip}}$. The $\alpha$ is a free parameter denoting the strength of the NRB, whose distribution is modeled using a third-order Chebyshev polynomial. The BKG denotes background contributions other than NRB. The known backgrounds are parameterized with the methods mentioned earlier, while the unknown non-$\SSB$ background is described by the normalized $\SSB$ sideband in data.

Assuming the interference between resonant and non-resonant processes is maximum in the $\SSB$ final state, the number of $\etac$ signal events is determined to be $786 \pm 42$ (destructive) and $358 \pm 35$ (constructive), where the uncertainties are statistical only. The corresponding fits are shown in Fig.~\ref{1sfit}.

The BF for $\etac\to\SSB$ is calculated by
\begin{footnotesize}
\begin{equation}\label{1sbf}
\textstyle{\displaystyle\mathcal{B}(\psip \to \gamma\etac)\cdot\displaystyle\mathcal{B}(\etac \to \SSB) = \frac{{N^{\rm obs}}} {\substack{N^{\rm tot}_{\psip} \cdot \prod_{i}\displaystyle\mathcal{B}_{i}\cdot\epsilon}}},
\end{equation}
\end{footnotesize}where $N^{\rm obs}$ is the number of signal events in data; $N^{\rm tot}_{\psip}$ is the total number
of $\psip$ events; $\prod_{i}\displaystyle\mathcal{B}_{i}$ is the product of the BFs of intermediate states, i.e., $\displaystyle\mathcal{B}(\Sigma^{0} \to \gamma\Lambda)\cdot\displaystyle\mathcal{B}(\bar{\Sigma^{0}}\to \gamma\bar{\Lambda})\cdot\displaystyle\mathcal{B}(\Lambda \to p\pi^{-})\cdot\displaystyle\mathcal{B}(\bar{\Lambda} \to \bar{p}\pi^{+})$, quoted from the PDG~\cite{Ref_pdg}; and $\epsilon$ is the detection efficiency, which is obtained from the signal MC events by applying the same selection criteria as for data. The product BF of
$\displaystyle\mathcal{B}(\psip \to \gamma\etac) \times\textstyle{\displaystyle\mathcal{B}}(\etac \to \SSB)$ is measured  for both the destructive solution and the constructive solution. The numerical results are listed in Table~\ref{list79}.

\begin{table}[htp]
\begin{center}
\caption{Results for the signal yields, the detection efficiencies, and the measured BFs for $\etac\to\SSB$ for destructive and constructive solutions. The first uncertainties are statistical and the second are systematic.}
\label{list79}
\scalebox{0.95}{
\begin{tabular}{ c c c c}
\hline
\hline
        \ \ \ \  \  Interference  \ \ \    \ \ \   &      \ \ \  $N^{\rm obs}$  \ \ \  \ \ \    &  \ \ \  $\epsilon~(\%)$  \ \ \   &  \  ${\cal B}$ ($\times 10^{-3}$) \\  \hline
        Destructive & 786 $\pm$ $42$ & 7.55 & 2.59 $\pm 0.14\pm 0.44$   \\ \hline
        Constructive & 358 $\pm$ $35$ & 7.55 & 1.18 $ \pm 0.12\pm 0.21$   \\
        \hline
        \hline
\end{tabular}
}
\end{center}
\end{table}

Detailed descriptions of the systematic uncertainties are provided in
the subsequent section.

\subsection{$\hc \to \SSB$}

\begin{figure}[htb]
\begin{center}
\includegraphics[width=.4\textwidth]{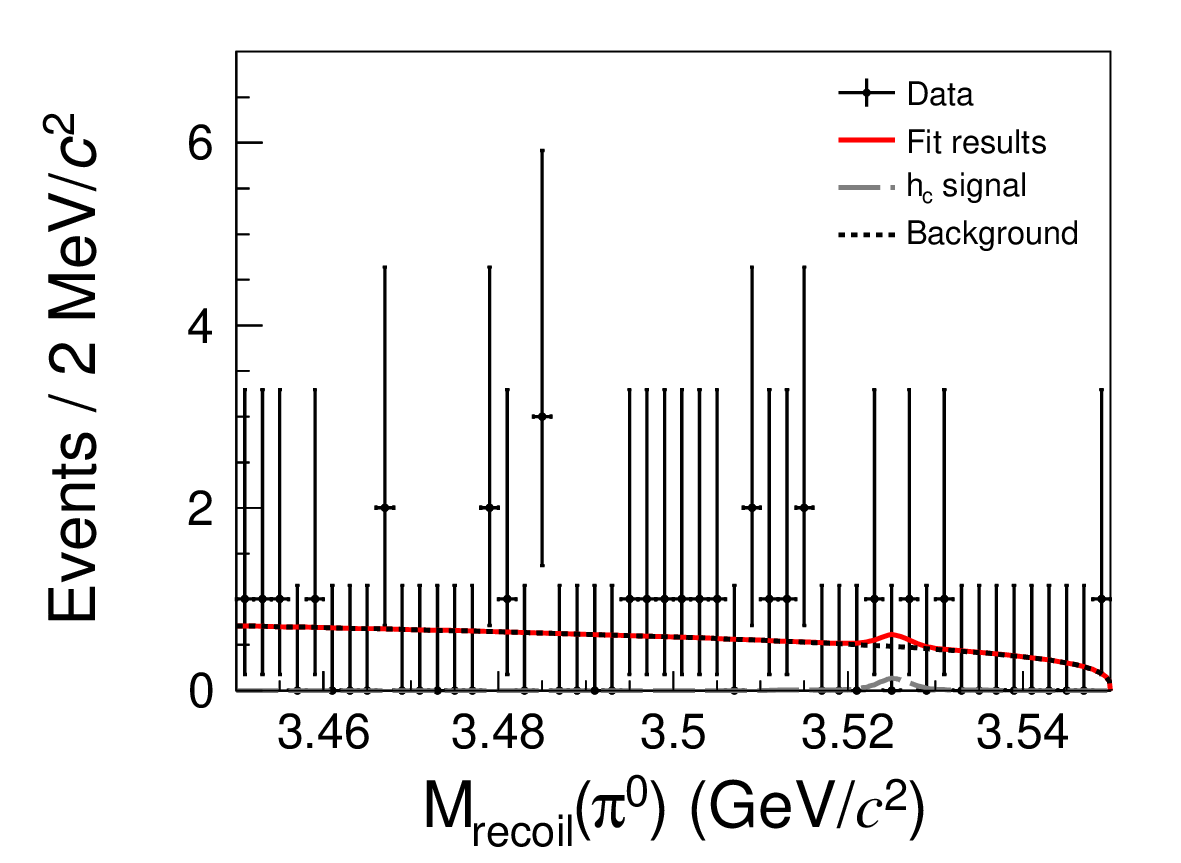}
\caption{The fit to the M$_{\rm recoil}(\pi^0)$ distribution of the accepted candidates for $h_c\to \Sigma^0\bar \Sigma^0$ in data.}
\label{hcfit}
\end{center}
\end{figure}

The signal yield of $\hc$ is extracted using an unbinned maximum likelihood fit to the $\pi^{0}$ recoil mass distribution, $M_{\rm recoil}(\pi^0)$, of the accepted candidates for $h_c\to \Sigma^0\bar \Sigma^0$ in data. The fit PDF is constructed as
\begin{equation}\label{hcsig}
N_{\mathrm{sig}} \cdot \text{MC shape} + N_{\mathrm{bkg}} \cdot \text{Argus}(m; r, s, p).
\end{equation}
The signal function is modeled by the MC-simulated shape, and the background function is parameterized with an ARGUS function, where the parameters $r$, $s$, and $p$ represent the mass threshold, the slope parameter, and the power, respectively~\cite{Ref_argus}, as shown in Fig.~\ref{hcfit}.

\begin{figure}[htb]
\begin{center}
\includegraphics[width=.4\textwidth]{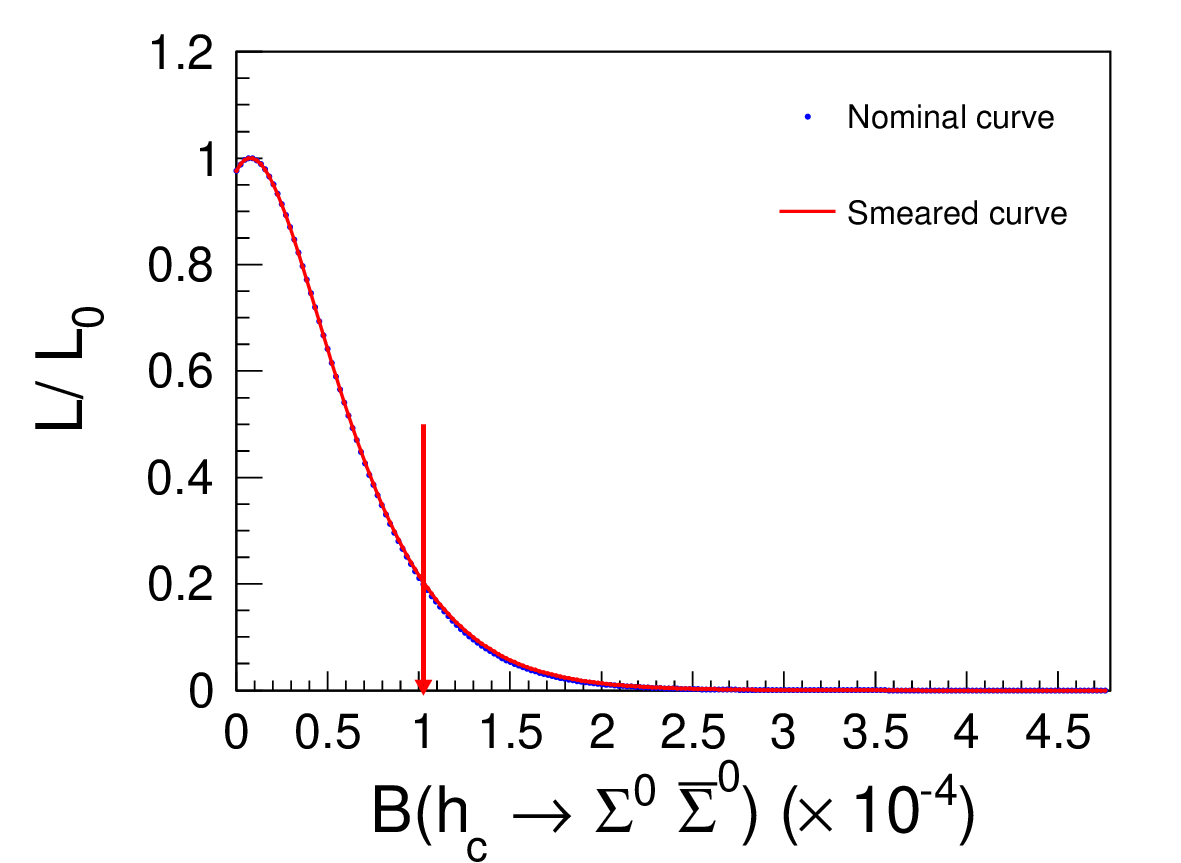}
\caption{The normalized likelihood distribution before~(blue dots) and after~(red line) convolving the multiplicative systematic uncertainties for $h_{c}\to \SSB$. The red arrow denotes the number of signal events corresponding to 90\% C. L. after being convolved with the multiplicative systematic uncertainties.}
\label{hcsmear}
\end{center}
\end{figure}

Since no obvious signal is observed, an upper limit on the BF of $\hc\to\SSB$ is determined at the 90\% confidence level (C.L.) using a Bayesian approach~\cite{Ref_bayes}, which is calculated by:
\begin{multline}\label{hcbf}
\textstyle{\displaystyle\mathcal{B}(h_{c} \to \SSB) < \frac{{N_{\rm obs}^{\rm UL}}} {\substack{N^{\rm tot}_{\psip} \cdot \displaystyle\mathcal{B}(\psip \to \pi^{0} h_{c}) \cdot\prod_{i}\displaystyle\mathcal{B}_{i}\cdot\epsilon}}}
\end{multline}
where $\prod_{i}\displaystyle\mathcal{B}_{i}$ is the product of the
BFs of intermediate states, i.e.,
$\displaystyle\mathcal{B}(\Sigma^{0} \to
\gamma\Lambda)\cdot\displaystyle\mathcal{B}(\bar{\Sigma^{0}}\to
\gamma\bar{\Lambda})\cdot\displaystyle\mathcal{B}(\Lambda \to
p\pi^{-})\cdot\displaystyle\mathcal{B}(\bar{\Lambda} \to
\bar{p}\pi^{+})\cdot\displaystyle\mathcal{B}(\pi^{0}\to\gamma\gamma)$,
quoted from the PDG~\cite{Ref_pdg}; and $\epsilon$ is the detection
efficiency obtained by the signal MC simulation, which is
5.39$\%$. The systematic uncertainty on the BF upper limit includes
contributions from both additive and multiplicative sources. The
additive sources comprise those from the signal shape and background
shape, both of which are considered separately to recalculate the
upper limit and the most conservative likelihood curve is taken. The
other sources listed in Table~\ref{su} are multiplicative. The total
multiplicative systematic uncertainty is then incorporated by
convolving the likelihood with a Gaussian
function~\cite{Ref_gauss}. The likelihood curves before and after
considering the multiplicative uncertainties are shown in
Fig.~\ref{hcsmear}.

The upper limit for BF of $\displaystyle\mathcal{B}(h_c \to \SSB)$ is determined to be $1.02\times10^{-4}$.

\section{Systematic Uncertainties}\label{sec:sysU}
The sources of systematic uncertainties are discussed below and the resulted numbers are listed in Table~\ref{su}. For each signal decay, the individual uncertainties are assumed to be independent and are added in quadrature to obtain the total systematic uncertainty.


\subsection{$\Lambda$($\bar{\Lambda})$ reconstruction}
The systematic uncertainty due to $\Lambda$($\bar{\Lambda})$ reconstruction, including proton/antiproton, charged pion tracking, PID, $\Lambda(\bar{\Lambda})$ decay length and mass window, is assigned to be $1.0\%$ for both $\Lambda$ and $\bar{\Lambda}$, by using the control sample of $J/\psi(\psi(3686))\to\Lambda\bar{\Lambda}$~\cite{Ref_llb}.

\subsection{Photon reconstruction}
The systematic uncertainty from photon reconstruction is assigned to be 0.5$\%$ for each photon, by using a control sample of $e^{+}e^{-} \to \gamma \mu^{+}\mu^{-}$~\cite{Ref_photon}.

\subsection{$\pi^{0}$ reconstruction}
The uncertainty due to $\pi^{0}$ reconstruction is conservatively estimated to be $1.0\%$, by using the control samples of $\psip \to \pi^{0}\pi^{0}\jpsi$ with $\jpsi \to l^{+}l^{-}$ and $e^{+}e^{-} \to \omega\pi^{0}$ at $\sqrt{s} = 3.773$ GeV.

\subsection{Kinematic fit}
The uncertainties in kinematic fit are estimated by comparing the difference of detection efficiency with and without the track helix parameters correction~\cite{Ref_kine}.

\subsection{$\pi^{0}$ mass window}
For the measurement of ${\cal B}(\etac \to \SSB)$, the uncertainty due to  $\pi^0$ mass window is estimated to be 1.4\% for destructive solution and 1.8\% for constructive solution by changing the position of mass window from the low and high sides by $\pm1.0 \sigma$ by 900 times~\cite{Ref_pi0}.

\subsection{$\Sigma^{0} \bar{\Sigma^{0}}$ mass window}
For the $\etac \to \SSB$ decay mode, the uncertainty  due to $\SSB$ mass window is determined to be 2.1$\%$ by employing the control sample of $\psip\to \pi^{0}\SSB$.

\subsection{MC statistics} The uncertainties due to MC statistics are assigned by $\sqrt{\frac{(1-\epsilon)}{N\epsilon}}$, where $\epsilon$ is the detection efficiency and $N$ is the total number of produced signal MC events.

\subsection{Mass fit}
For the $\etac \to \SSB$ decay mode, the uncertainties caused by mass fit include the components listed below.

\noindent\hspace{2em}{\itshape Damping function}: The systematic uncertainty in the damping function is estimated by changing the KEDR form $\frac{E_{0}^{2} }{E_{\gamma }E_{0}+(E_{\gamma }-E_{0})^{2}}$~\cite{Ref_kedr} to the CLEO form written as $e^{-\frac{E_{\gamma}^{2}}{8\beta^{2}}}$~\cite{Ref_cleo}.

\noindent\hspace{2em}{\itshape NRB shape}: The uncertainty from NRB shape is estimated by varying the parameters of the PDF, modeled by a third-order Chebyshev polynomial, from fixed to floating values.

\noindent\hspace{2em}{\itshape Sideband region}: The systematic uncertainty associated with the 2D $\SSB$ sideband background is estimated by zooming in and out the sideband regions by 10 MeV/$c^{2}$ around the $\SSB$ mass window.

\noindent\hspace{2em}{\itshape Mass and width of $\etac$}: The uncertainty caused by the $\etac$ mass and width is determined by changing the nominal value by one standard deviation. The difference relative to the nominal result is taken as the systematic uncertainty.

For the $\hc \to \SSB$ decay mode, the uncertainties caused by mass fit include the sources listed below.

\noindent\hspace{2em}{\itshape Signal shape}: The uncertainty from signal shape is estimated by convolving the MC-determined shape with a Gaussian function characterized by free parameters. The difference in the final results is taken as the systematic uncertainty.

\noindent\hspace{2em}{\itshape Background shape}: To estimate the uncertainty associated with the description of the background, its shape in the fit is changed from the Argus function to a second-order polynomial function with free parameters. The difference in the result is assigned as the systematic uncertainty.

\subsection{MC model}
For the $\hc\to\SSB$ decay mode, to estimate the uncertainty of the MC mode for the $\hc\to\SSB$ decay mode, we set the angular distribution of $\Sigma^0$ in the $\hc$ rest frame to be $1+\cos^{2} \theta/1-\cos^{2} \theta$. The largest efficiency difference relative to the PHSP model is taken as the systematic uncertainty.

\subsection{Quoted BFs}
The uncertainties arising from the quoted BFs are taken from the PDG~\cite{Ref_pdg}, which are 13.9\%, 6.8\%, and 0.8\% for $\psi(3686)\to \gamma \eta_c$, $\psi(3686)\to \pi^0 h_c$, and $\Lambda\to p\pi^-$, respectively; while negligible for both $\pi^0\to \gamma\gamma$ and $\Sigma^0\to \gamma \Lambda$.


\subsection{The total number of $\psi(3686)$ events}
The uncertainty in the total number of $\psi(3686)$ events is 0.5$\%$~\cite{Ref_total}.\newline

\begin{table*}[htp]
\begin{center}
\caption{Relative systematic uncertainties (in \%) for the BF measurements of ${\cal B}(\etac \to \SSB)$ and $\displaystyle\mathcal{B}(h_{c} \to \SSB)$. The system uncertainties in parentheses are the ones without the relative uncertainties from ${\cal B}(\psip \to \gamma\etac)$ and ${\cal B}(\psip \to \pi^0\hc)$.}
\label{su}
\scalebox{0.95}{
\begin{tabular}{l c c c c c c}
\hline
\hline
	Source & $\displaystyle\mathcal{B}(\etac \to \SSB)$(destructive)  \ \ \ &  \ \ \  \ \ \ $\displaystyle\mathcal{B}(\etac \to \SSB)$(constructive) \ \ \ \ \ \  &  \ \ \ $\displaystyle\mathcal{B}(\hc \to \SSB)$ \\   \hline
 	$\Lambda(\bar{\Lambda})$ reconstruction  &   2.0 &  2.0&   2.0    \\
	Photon reconstruction  &  1.5  & 1.5 & 1.0\\
    $\pi^{0}$ reconstruction  &  - &  - &  1.0  \\
	Kinematic fit   &  6.9  & 6.9 &  3.8\\
    $\pi^{0}$ mass window  &   1.4  &  1.8  &  -\\
	$\SSB$ mass window  &   2.1 &  2.1  &  2.1   \\
 	MC statistics   & 0.4  &  0.4   & 0.3  \\
	Damping function  &   3.0  & 3.1 &  - \\
    NRB shape  &   2.1  &  5.6  &  -\\
	$N_{\pi^{0}\SSB}$  &  Negligible  & Negligible  &  -\\
	Sideband region  &    4.0  &  5.1  &  - \\
	Mass of $\etac$  &   0.2  &  0.1  &  -  \\
	Width of $\etac$  &   0.9  & 0.4 &  - \\
    Signal shape  &   -  & - &  Negligible \\
    Background  &   -  & - &  2.2 \\
    MC model     &  - &  - & 2.3\\
    Quoted BFs  &   14.0  &   14.0 &   6.9\\
	Total number of $\psip$ events &   0.5  &   0.5  &   0.5\\
    \hline
	Sum &  16.9(9.7) &   18.0(11.5)  &  9.1(5.9)  \\
 \hline
 \hline
\end{tabular}
}
\end{center}
\end{table*}

\section{Summary}

Using $(2712.4 \pm 14.3) \times 10^6~\psip$ events collected by the BESIII detector, the $\etac \to \SSB$ decay is observed for the first time. Its BF is measured to be $\mathcal{B}(\etac \to \SSB)= (2.59 \pm 0.14(stat) \pm 0.44(syst)) \times 10^{-3}$ for destructive interference and $(1.18 \pm 0.12(stat) \pm 0.21(syst)) \times 10^{-3}$ for constructive interference. In addition, we search for the decay $h_c \to \SSB$ for the first time and no significant signal is observed. The upper limit of its BF is determined to be $\mathcal{B}(h_{c} \to \SSB)<1.02\times 10^{-4}$ at the 90\% C.L. The obtained results are inconsistent with theoretical calculations, which give $(4.82\sim9.56)\times10^{-4}$ for $\mathcal{B}(\etac \to \SSB)$ and $(5.57\sim7.08)\times10^{-4}$ for $\mathcal{B}(\hc \to \SSB)$~\cite{Ref_bb}, assuming the isospin conservation between charged and neutral modes.


\section{ACKNOWLEDGMENTS}


The BESIII Collaboration thanks the staff of BEPCII (https://cstr.cn/31109.02.BEPC) and the IHEP computing center for their strong support. This work is supported in part by National Key R\&D Program of China under Contracts Nos. 2023YFA1606000, 2023YFA1606704, 2025YFA1613900; National Natural Science Foundation of China (NSFC) under Contracts Nos. 11635010, 11935015, 11935016, 11935018, 12025502, 12035009, 12035013, 12061131003, 12192260, 12192261, 12192262, 12192263, 12192264, 12192265, 12221005, 12225509, 12235017, 12342502, 12361141819, 12535005; the Chinese Academy of Sciences (CAS) Large-Scale Scientific Facility Program; the Strategic Priority Research Program of Chinese Academy of Sciences under Contract No. XDA0480600; CAS under Contract No. YSBR-101; 100 Talents Program of CAS; The Institute of Nuclear and Particle Physics (INPAC) and Shanghai Key Laboratory for Particle Physics and Cosmology; Agencia Nacional de Investigación y Desarrollo de Chile (ANID), Chile under Contract No. ANID CCTVal CIA250027; ERC under Contract No. 758462; German Research Foundation DFG under Contract No. FOR5327; Istituto Nazionale di Fisica Nucleare, Italy; Knut and Alice Wallenberg Foundation under Contracts Nos. 2021.0174, 2021.0299, 2023.0315; Ministry of Development of Turkey under Contract No. DPT2006K-120470; National Research Foundation of Korea under Contract No. RS-2026-25486791; National Science and Technology fund of Mongolia; Polish National Science Centre under Contract No. 2024/53/B/ST2/00975; STFC (United Kingdom); Swedish Research Council under Contract No. 2019.04595; U. S. Department of Energy under Contract No. DE-FG02-05ER41374




\end{document}